\documentclass[useAMS,usenatbib,usegraphicx]{mn2e}

\title[Kinematics of OB-associations] {Kinematics of OB-associations and the new
reduction of the Hipparcos data} \author[A.M. Mel'nik and A.K.
Dambis]{A.M.Mel'nik\thanks{E-mail: anna@sai.msu.ru}   and A.K.
Dambis\\ Sternberg Astronomical Institute, Universitetskii
pr. 13, Moscow, 119992 Russia}

\begin{document}

\date{Accepted 2009 December 00. Received 2009 December 00; in original form 2009 December 00}


\maketitle

\label{firstpage}

\begin{abstract}
The proper motions of OB-associations computed using the old
\citep{hipparcos1997} and new \citep{vanleeuwen2007} reductions of
the Hipparcos data are in a good agreement with each other. The
Galactic rotation curve derived from an analysis of line-of-sight
velocities and proper motions of OB-associations is almost flat in
the 3-kpc neighborhood of the Sun. The angular rotation velocity
at the solar distance is $\Omega_0=31\pm1$ km s$^{-1}$ kpc$^{-1}$.
The standard deviation of the velocities of OB-associations from
the rotation curve is $\sigma=7.2$ km s$^{-1}$. The distance scale
for OB associations \citep{blahahumphreys1989} should be shortened
by 10--20\%. The residual velocities of OB-associations calculated
for the new and old reductions differ, on average, by 3.5 km
s$^{-1}$. The mean residual velocities of OB-associations in the
stellar-gas complexes depend only slightly on the data reduction
employed.

\end{abstract}

\begin{keywords}
Galaxy: kinematics and dynamics -- open clusters and associations
\end{keywords}

\section{Introduction}

The original data from the Hipparcos satellite were obtained in
the form of time moments for  transitions of stars through the
diffraction grating which  were then transformed  into the angular
positions along the scan direction \citep{kovalevsky2002}. A
better understanding of the peculiarities in the rotation of the
satellite and increased computer power allowed
\citet{vanleeuwen2007} to abandon the intermediate reduction on
the great-circle, which was one of the sources of noise. A
global iterative solution resulted in a factor of four improvement
in the accuracy of the astrometric data  for bright stars
($m_V<8^m$) \citep{vanleeuwen2007}.

In this paper we compare the proper motions, parallaxes, and
residual velocities of OB-associations derived for the old
\citep{hipparcos1997} and new \citep{vanleeuwen2007}  reductions
of the Hipparcos data. We also determine the parameters of the
Galactic rotation curve and  improve the distance scale for
OB-associations. A good agreement between the old and new results
is indicative of high quality of the astrometric data obtained for
bright stars.

OB-associations are large groups of high-luminosity stars, though
the sky-plane size of most of them does not exceed 300 pc. They
often have several centers of concentration. The catalog of
\citet{blahahumphreys1989} of stars of OB associations includes 91
groups located within 3.5 kpc from the Sun.  There are several
partitions of high-luminosity stars (OB-stars and red supergiants)
into OB-associations \citep{blahahumphreys1989, garmany1992,
melnik1995}, but all partitions used the photometric data obtained
by \citet{blahahumphreys1989}. We don't use the partition by
\citet{garmany1992} because it covers only a restricted interval
of Galactic longitudes 55-150$^\circ$. Our own partition
\citep{melnik1995} is also unsuitable for kinematical studies: the
compact centers of OB-associations derived by using cluster
analysis method include only few stars with known kinematical
data. All associations considered here are unbound objects. The
new partition \citep{melnik1995} clearly shows that even compact
centers of OB-associations have very large velocity dispersions to
be bound objects.

\citet{blahahumphreys1989} derived distances for OB-associations
by averaging  distance moduli of individual stars. Membership in
the associations is based on positions, spectral type, luminosity,
and resulting photometric distance. The above authors used  their
own calibration which differ little from the previously published
ones \citep [see comments in][]{humphreys1984}.

The mass measurements of proper motions of blue stars became
available due to the inclusion of a list of OB-stars into the
Hipparcos input  catalog  \citep{dezeeuw1999}. The mean visual
magnitude of the stars of OB-associations with known proper
motions is $m_V=7.3^m$ and $m_V=7.8^m$ in the areas $0<r<3.5$ kpc
and $1.5<r<3.5$ kpc, respectively. Thus they are bright enough,
and we can expect a conspicuous increase in the accuracy of their
astrometric data.

\section{Results}

\subsection{Proper motions of OB-associations}

\begin{figure*}
\resizebox{\hsize}{!}{\includegraphics{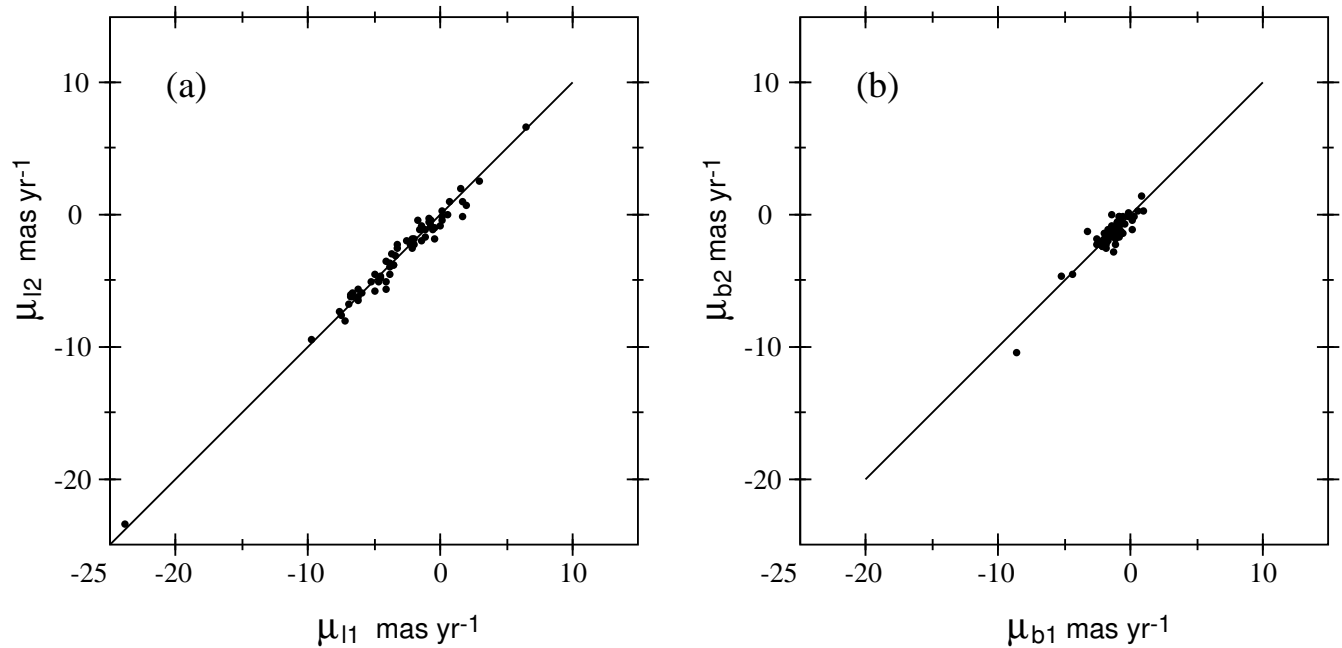}} \caption{ The
proper motions of 64 OB-associations derived for the old (index
$1$) and new (index $2$) reductions of the Hipparcos data: (a)
proper motions along $l$-coordinate, (b) those along
$b$-coordinate. The line goes at the angle of 45$^\circ$.}
\label{compare-mu}
\end{figure*}

  The OB-associations have quite reliable distances, which are
accurate, on the average, to  6\%  without the allowance for the
uncertainly of the zero-point of the distance scale (see
Appendix). Table~1, which is available in electronic form, gives
the line-of-sight velocities and proper motions of OB-associations
based on kinematical data of individual stars. We used the median
estimations instead of mean values for velocities to reduce the
influence of stars which velocities deviate strongly from the mean
value. For every OB-association from the catalog by
\citet{blahahumphreys1989} we give the mean galactic coordinates
$l$ and $b$; the mean heliocentric distance $r$; median
line-of-sight velocity $V_r$; the dispersion of line-of-sight
velocities $\sigma_{vr}$,  and number of stars $n_{vr}$ with known
line-of-sight velocity. We adopt the line-of-sight velocities from
the catalog by \citet{barbierbrossat2000}. We use only the
velocities measured with errors of less than 10 km s$^{-1}$, which
corresponds to the quality estimations A, B, and C. Table~1 also
gives the median proper motions of OB-associations along $l$- and
$b$-coordinates, $\mu_l$ and $\mu_b$. The data obtained for the
reduction of 1997 \citep{hipparcos1997} are denoted by subscript
1, whereas those based on the reduction by \citet{vanleeuwen2007}
are marked by subscript 2. For each OB-association we give the
dispersions of proper motions, $\sigma_{\mu l}$ and $\sigma_{\mu
b}$, as well as a number of stars $n_\mu$ with known proper
motion. The last column gives the total number of stars with known
photometric measurements, $N_t$, used to determine the distances
for OB-associations. The distances listed in Table~1 agree with
the short distance scale for classical Cepheids
\citep{berdnikov2000}. They are equal to the distances from the
catalog by \citet{blahahumphreys1989}, $r_{BH}$, multiplied by  a
factor of 0.8, $r=0.8 r_{BH}$ \citep{sitnik1996,dambis2001}.

We determine the median proper motions for 64 OB-associations
containing at least two stars with known proper motions and the
median line-of-sight velocities for 70 OB-associations containing
at least two stars with known line-of-sight velocities. The
velocity of each OB-association is based, on average, on 12
line-of-sight velocities and 11 proper motions of individual
stars.

Let us consider a sample of 64 OB associations containing at least
two stars with known proper motions. Fig.~\ref{compare-mu} shows
the proper motions of OB-associations computed using the old and
new reduction of the Hipparcos catalog. It is immediately apparent
that the new reduction doesn't bring dramatic changes into proper
motions of OB-associations. The root-mean-square (rms) difference
between the proper motions $\mu_{l1}$ and $\mu_{l2}$ derived for
the old and new reductions is $\Delta_{\mu_l}=0.58$ mas yr$^{-1}$,
and the rms difference  of the proper motions $\mu_{b1}$ and
$\mu_{b2}$ is $\Delta_{\mu_b}=0.62$ mas yr$^{-1}$. These
deviations in proper motions translate into tangential-velocity
deviations of $\Delta_{tang}=4.6$ km s$^{-1}$. That doesn't exceed
the standard deviation of the velocities of OB-associations from
rotation curve.

Note that the values given here are very much sample dependent.
For example, the deviations calculated for a sample of 29
OB-associations with proper motions based on at least 10
individual stars reduce  to $\Delta_{\mu_l}=0.41$,
$\Delta_{\mu_b}=0.50$ mas yr$^{-1}$ and $\Delta_{tang}=2.7$ km
s$^{-1}$.

\subsection{Dispersion of stellar proper motions  inside OB-associations}

\begin{figure*}
 \resizebox{\hsize}{!}{\includegraphics{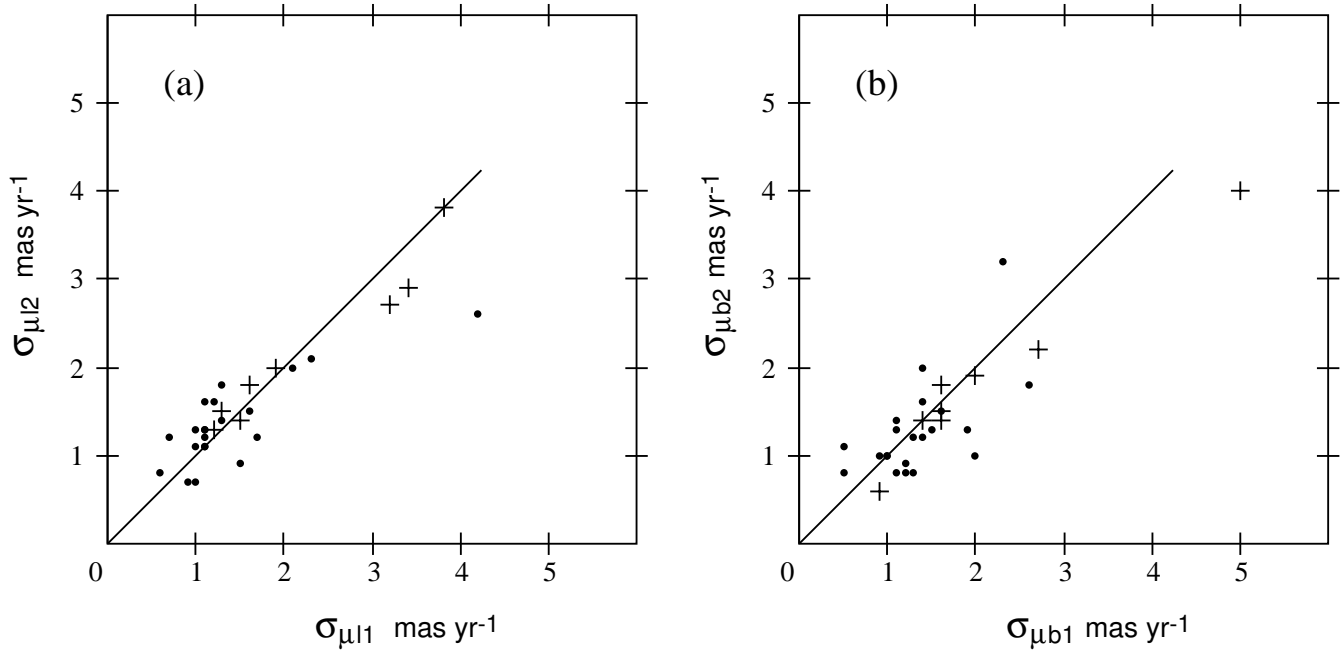}}
\caption{The dispersions of the observed proper motions for 29
OB-associations obtained for the old (index $1$) and new (index
$2$) reductions of the Hipparcos data: (a) proper motions along
$l$-coordinate, (b) those along $b$-coordinate. Associations
located within 1 kpc are indicated by pluses, while those located
at the distances $r>1$ kpc are shown by dots. The line goes at the
angle of 45$^\circ$.} \label{compare-sigma}
\end{figure*}

Let us consider the dispersions of individual stellar proper
motions, $\sigma_{\mu l}$ and  $\sigma_{\mu b}$,  in 29
OB-associations with at least 10 Hipparcos stars.
Fig.~\ref{compare-sigma} compares these dispersions computed for
the old and new reductions of the Hipparcos data. It is obvious
that closer OB-associations have larger  proper motions (in
absolute value) and, consequently, greater $\sigma_{\mu_l}$ and
$\sigma_{\mu_b}$. The pluses  and dots  show the associations
lying within 1 kpc and beyond 1 kpc from the Sun, respectively.
Generally, we see  a weak  decrease in the dispersions of proper
motions in the new reduction. The  mean  dispersion in the plane
of the sky decreases  to 1.67 mas yr$^{-1}$ in the new reduction
against 1.81 mas yr$^{-1}$ in the old one. But the corresponding
velocity dispersion remains practically at the same level,
$\sigma_{tang}=9.7$ km s$^{-1}$. For comparison, the mean
dispersion of line-of-sight velocities computed for 28
OB-associations containing at least 10 stars with known
line-of-sight velocities amounts $\sigma_{sight}=12.3$ km
s$^{-1}$.

One value of $\sigma_{\mu l}$  decreases   from $4.2$ mas
yr$^{-1}$ in the old reduction to $2.6$ mas yr$^{-1}$ in the new
reduction (Fig.~\ref{compare-sigma}a). This significant change occurs
in the OB-associations Ara OB1A ($r=1.1$ kpc). The decrease is due
to changes in the proper motions of the stars HD 150135 and
150136.

\subsection{Trigonometric parallaxes of OB-associations}

We use  van Leeuwen's data to calculate the trigonometric
parallaxes $p_t$ for OB-associations as the median values of the
parallaxes of individual stars. We derive them for 29
OB-associations containing at least 10 stars with known
parallaxes. This allowed us to compare the distance scale for
OB-associations based on photometric distances by
\citet{blahahumphreys1989}, $r_{BH}$, with  that based on the
Hipparcos data. We converted the photometric distances, $r_{BH}$,
into photometric parallaxes, $p_{BH}=1/r_{BH}$, and made a
least-square solution for a set of 29 linear equations,
$p_t=k'p_{BH}$. The coefficient $k'$ is equal to $k'=1.14\pm0.06$
and hence coincides with the value derived for the old reduction
of the Hipparcos data \citep{dambis2001}. The standard deviation
of the association parallaxes from the derived relationship is
$\sigma=0.44$ mas. The distance-scale factor $k$ relating
photometric $r_{BH}$ and parallax-based distances, $r=kr_{BH}$, is
$k=0.88\pm0.05$. The exclusion of the nearest OB-association Sco
OB2 (r=0.1 kpc) changes the coefficient $k'$ to $k'=1.27\pm0.09$,
which corresponds to a distance-scale factor of $k=0.78\pm0.07$.

In order to increase the sample size, we turned to the second part
of the high-luminosity-star catalog of \citet{blahahumphreys1989},
which includes field stars. The distances to these stars were
determined using the same photometric calibration as was used for
stars of OB-associations. For 1006 field stars we found
trigonometric parallaxes $p_t$ determined with the errors of less
than 1 mas. Note, that 534 of them are located within 1 kpc. The
coefficient in the relation between trigonometric and photometric
parallaxes is $k'=1.20\pm0.02$. The rms deviation of the stellar
parallaxes from the relation derived is $\sigma=1.5$ mas. The
distance-scale coefficient is $k=0.83\pm0.02$ and practically
coincides with the value calculated for the old reduction,
$k=0.84\pm0.02$ \citep{dambis2001}. Thus the old and new
reductions of the Hipparcos data yield very similar results
indicating that the distance scale for OB-associations by
\citet{blahahumphreys1989} should be shrunk  by
$10\textrm{--}20$\%.

\addtocounter{table}{1}

\begin{table*}
\caption{Parameters of the rotation curve}
 \begin{tabular}{lccccccccl}
  \hline
 $R_0$ & $\Omega_0$ & $\Omega'_0$ & $\Omega''_0$ & $u_0$ & $v_0$ & $k$ & A & $\sigma$ & $\chi^2$  \\
  kpc &  km s$^{-1}$  & km s$^{-1}$  & km s$^{-1}$  &
  km s$^{-1}$ & km s$^{-1}$ & &km s$^{-1}$ &km s$^{-1}$ &  \\
    & kpc$^{-1}$ & kpc$^{-2}$ & kpc$^{-3}$ &
   & & & kpc$^{-1}$& &  \\
  \\[-7pt] \hline\\[-7pt]
7.1  & 30.7 & -5.1 & 1.6 & 7.6 & 11.4 & 0.77& 17.9 & 7.1990 & 134.3\\
7.5  & 30.6 & -4.7 & 1.4 & 7.7 & 11.6 & 0.78& 17.7 & 7.1693 & 133.2\\
8.0  & 30.4 & -4.4 & 1.3 & 7.9 & 11.8 & 0.79& 17.6 & 7.1407 & 132.2\\
8.5  & 30.4 & -4.1 & 1.1 & 7.9 & 12.0 & 0.79& 17.5 & 7.1194 & 131.4\\
9.0  & 30.3 & -3.8 & 1.0 & 8.1 & 12.2 & 0.80& 17.3 & 7.1045 & 130.8\\
Err  & $\pm0.9$ & $\pm0.2$ & $\pm0.2$ & $\pm1.0$ & $\pm1.3$ & $\pm0.1$ & $\pm0.8$ &&\\
 \hline
\end{tabular}
\end{table*}

\begin{table*}
  \caption{Mean  residual velocities of OB-associations in the stellar-gas complexes}
  \begin{tabular}{lcccccl}
  \hline
  Region& {\it R}, kpc & $V_R$, & $V_{\theta}$,  & {\it l}, deg. & {\it r}, kpc & Associations \\
    &  & km s$^{-1}$ & km s$^{-1}$  &  &  &  \\
  \\[-7pt]\hline\\[-7pt]
  Sagittarius & 5.6 & $+9.9\pm2.4$ & $-1.0\pm1.9$ & 8--23  & 1.3--1.9 & Sgr OB1, OB7, OB4, Ser OB1, OB2, \\
  & & & & & &  Sct OB2, OB3;\\
  Carina & 6.5 & $-5.8\pm3.3$ & $+4.7\pm2.2$ & 286--315  & 1.5--2.1 & Car OB1, OB2, Cru OB1, Cen OB1,\\
  & & & & & &   Coll 228, Tr 16, Hogg 16, NGC 3766, 5606;\\
  Cygnus & 6.9 & $-5.0\pm2.6$ & $-10.4\pm1.4$ & 73--78  & 1.0--1.8 & Cyg OB1, OB3, OB8, OB9; \\
  Local System & 7.4 & $+5.3\pm2.8$ & $+0.6\pm2.5$ & 0--360  & 0.1--0.6 & Per OB2, Mon OB1, Ori OB1, Vela OB2, \\
  & & & & & &   Coll 121, 140, Sco OB2; \\
  Perseus & 8.4 & $-6.7\pm3.0$ & $-5.9\pm1.5$ & 104--135  & 1.8--2.8 & Per OB1, NGC 457, Cas OB8, OB7, OB6, \\
  & & & & & &  OB5, OB4, OB2, OB1, Cep OB1;\\
  \hline
\end{tabular}
\end{table*}

\subsection{Rotation curve}

An analysis of the line-of-sight velocities and proper motions of
OB-associations derived for the old reduction of the Hipparcos
data suggests that Galactic rotation curve is practically flat and
that it is characterized by a large angular rotation velocity at
the solar distance, $\Omega_0=31\pm1$ km s$^{-1}$ kpc$^{-1}$
\citep{dambis2001,melnik2001}. The large value of $\Omega_0$ was
also derived from an analysis of the kinematics of blue
supergiants, $\Omega_0=29.6\pm1.6$ km s$^{-1}$ kpc$^{-1}$
\citep{zabolotskikh2002}, and from the kinematics of star-forming
regions with the proper motions and parallaxes based on the VLBA
measurements, $\Omega_0=30\pm1$ \citep{reid2009}. In this paper we
determine the parameters of the rotation curve from an analysis of
the kinematics of OB-associations based on the new reduction of
the Hipparcos data.

Let us suppose that the motion of young objects of the disk obeys
a circular rotation law. We can then  write the
equations for the line-of-sight velocities and proper motions in
the following form:

\begin{equation}
\begin{array}{l}

V_r=R_0(\Omega-\Omega_0)\sin l \cos b  \\
\qquad -(u_0 \cos l \cos b+v_0\sin l \cos b+w_0 \sin b),
\end{array}
\end{equation}

\begin{equation}
\begin{array}{l}
4.74\mu_l r=R_0(\Omega-\Omega_0)\cos l - \Omega r \cos b \\
\qquad -(-u_0\sin l +v_0\cos l).
\end{array}
\end{equation}

These are the so-called Bottlinger equations, where $\Omega$ and
$\Omega_0$ are the angular rotation velocities calculated at the
Galactocentric distance $R$ and and at the distance of the Sun
$R_0$, respectively. The velocities $u_0$ and $v_0$ characterize
the solar motion with respect to the centroid of objects
considered in the direction toward the Galactic center and in the
direction of galactic rotation, correspondingly. The velocity
$w_0$ is directed along the $z$-coordinate and we set it equal to
$w_0=7.5$ km s$^{-1}$ (see section 2.5).  The factor 4.74 converts
the left-hand part of the equation (2) (where proper motion is in
mas yr$^{-1}$ and the distance is in kpc) into the units of km
s$^{-1}$.

 We expanded the angular rotation velocity $\Omega$  at
Galactocentric distance $R$ into a power series in $(R-R_0)$:

\begin{equation}
 \Omega= \Omega_0 + \Omega'_0(R-R_0) + 0.5\Omega''_0(R-R_0)^2,
\end{equation}

\noindent where $\Omega'_0$ and $\Omega''_0$ are its  first and
second derivatives taken at the solar Galactocentric distance.

We solve the  equations for the line-of-sight velocities and
proper motions jointly and use weight factors $p_{vr}$ and
$p_{vl}$ to allow for observational errors and "cosmic" velocity
dispersion:

\begin{equation}
p_{vr} =(\sigma_0^2+\varepsilon^2_{vr})^{-1/2},
\end{equation}

\begin{equation}
p_{vl} =(\sigma_0^2+(4.74\varepsilon_{\mu l}r)^2)^{-1/2},
\end{equation}

We adopt $\sigma_0=7.0$ km s$^{-1}$, which is approximately equal
to the rms deviation of the velocities from the rotation curve
\citep[for more details see][]{dambis1995, dambis2001}.

 The  errors of the median velocities $\varepsilon_{vr}$ and
$\varepsilon_{vl}$ depend on the velocity dispersion and number of
stars with known kinematics in the association:

\begin{equation}
\varepsilon_{vr}=\frac{\sigma_{vr}}{\sqrt{n_{vr}}}
\end{equation}

\begin{equation}
\varepsilon_{vl}=\frac{4.74r\sigma_{\mu l}}{\sqrt{n_\mu}}.
\end{equation}

Our approach includes two scaling parameters: the  solar
Galactocentric distance $R_0$ and the distance-scale coefficient
$k$, $r=kr_{BH}$. There is currently no consensus of opinion on
the exact value of $R_0$:  different authors report values in the
range $R_0=7.0-9.0$ \citep[see, e.g., a review
by][]{nikiforov2004}. For this reason, we calculated the
parameters of the rotation curve and the distance-scale
coefficient for five  values: $R_0=7.1$, 7.5, 8.0, 8.5, 9.0 kpc.
In our previous works we adopted $R_0=7.1$ \citep{dambis2001,
melnik2001}, which we derived from an analysis of Cepheid
line-of-sight velocities \citep{dambis1995} and from the spatial
distribution of globular clusters \citep{rastorguev1994}.

To determine the rotation curve, we selected 70 line-of-sight
velocities and 62 proper motions of OB-associations based on at
least two stars. We discarded the proper motions for two distant
OB-associations R 103 ($r=3.2$ kpc) and Ara OB1B ($r=2.8$ kpc)
having inexplicably large residual velocities along the
$z$-coordinate, $V_z=-31$ and $V_z=-24$ km s$^{-1}$. Note that the
old reduction of the Hipparcos catalog also yields large $V_z$
velocity components for these associations, $V_z\approx-22$ km
s$^{-1}$.

We use standard least square method \citep{press1987} to solve the
system of 132 equations, which are linear in the parameters
$\Omega_0$, $\Omega'_0$, $\Omega''_0$, $u_0$, and $v_0$,   for
each value of non-linear parameter $k$. We then determine the
value of $k$ that minimizes the sum of squared normalized residual
velocities $\chi^2$ -- which has a unique minimum in the interval
$(0.4,1.2)$ -- and estimate its standard error using the technique
proposed by \citet{hawley1986}.

Table 2 lists the parameters $\Omega_0$, $\Omega'_0$,
$\Omega''_0$, $u_0$, $v_0$, and $k$ calculated for  different
values of $R_0$. It also gives the rms residual $\sigma$, the
value of $\chi^2$, and the standard errors of the parameters. It
clearly shows that $\Omega_0$, $u_0$, $v_0$, and $k$ are
practically independent of the choice of $R_0$. The parameters
$\Omega'_0$ and $\Omega''_0$ vary conspicuously with $R_0$,
whereas the Oort constant $A=-0.5R_0\Omega'$ remains practically
unchanged, $A=17.3\textrm{--}17.9$ km s$^{-1}$ kpc$^{-1}$.
Generally, we can observe a weak tendency for decreasing  $\chi^2$
with increasing $R_0$. Note, however, that the circular rotation
law is too simplified to describe the motion of OB-associations
(see section 2.6).   The kinematical parameters  derived for the
old and new reductions of the Hipparcos catalog agree  within the
errors (compare Table 4 in \citet{dambis2001} with Table 2 in this
paper). We found  the velocity field of OB-associations to be very
robust, so that the exclusion of objects with velocity components
derived from 2-4 stars has no significant effect on the results.

\begin{figure}
 \resizebox{\hsize}{!}{\includegraphics{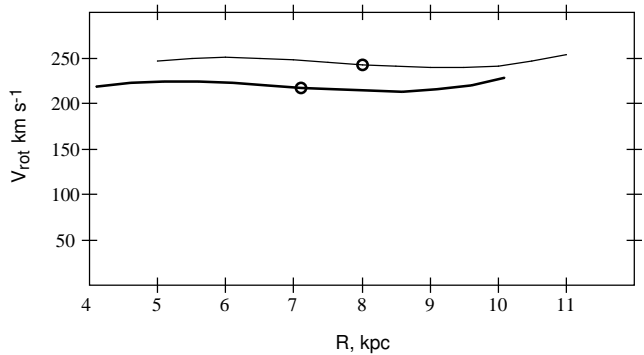}}
\caption{  The Galactic rotation curve derived from an analysis of
line-of-sight velocities and proper motions of OB-associations for
the adopted solar Galactocentric distances of $R_0=7.1$ and
$R_0=8.0$ kpc. The position of the Sun is shown by a circle.}
\label{rotation-curve}
\end{figure}

Fig.~\ref{rotation-curve} shows the Galactic rotation curves
within the 3 kpc of the Sun computed for $R_0=7.1$ kpc and
$R_0=8.0$ kpc. It is immediately apparent that the rotation curve
is practically flat in both cases. The linear velocity at the
solar distance is equal to $\Theta_0=218$ and $\Theta_0=243$ km
s$^{-1}$, respectively and remains constant within 3 kpc from the
Sun with the accuracy $\pm3$\%.

\subsection{Motions along the $z$-coordinate}

Let us consider motions along the $z$-coordinate. The velocities
of OB-associations, $V_z$, calculated from proper motions $\mu_b$
and line-of-sight velocities $V_r$:

\begin{equation}
V_z =4.74 r \mu_b \cos b + V_r \sin b,
\end{equation}

\noindent determine the solar velocity  $w_0$. Its value derived
for 59 OB-associations is $w_0=7.6\pm0.7$ ($\sigma_{vz}=5.2$ km
s$^{-1}$) and $w_0=7.5\pm0.6$ ($\sigma_{vz}=5.0$ km s$^{-1}$) for
the old and new reductions, respectively.

\subsection{Residual velocities}

Residual velocities characterize  non-circular motions in the
Galactic disk. They are determined as differences between the
observed velocities and model velocities computed in terms of the
circular rotation law with adopted components of the solar motion.
We calculated the distribution of the residual velocities of
OB-associations in the Galactic plane for the different values of
$R_0$. For this aim we selected 59 OB-associations containing at
least two stars with known line-of-sight velocities and proper
motions using the new reduction of the Hipparcos data. We set the
distance-scale coefficient equal to $k=0.80$ in all cases and
adopt the values of the  other parameters from the corresponding
lines of Table 2. The fields of residual velocities are nearly the
same for different $R_0$ and we therefore show only the field
computed for $R_0=7.5$ kpc (Fig.~4). Note that the residual
velocities calculated for $R_0=7.1$ and $R_0=9.0$ kpc differ, on
average, 1.1 km s$^{-1}$.

\begin{figure}
 \resizebox{\hsize}{!}{\includegraphics{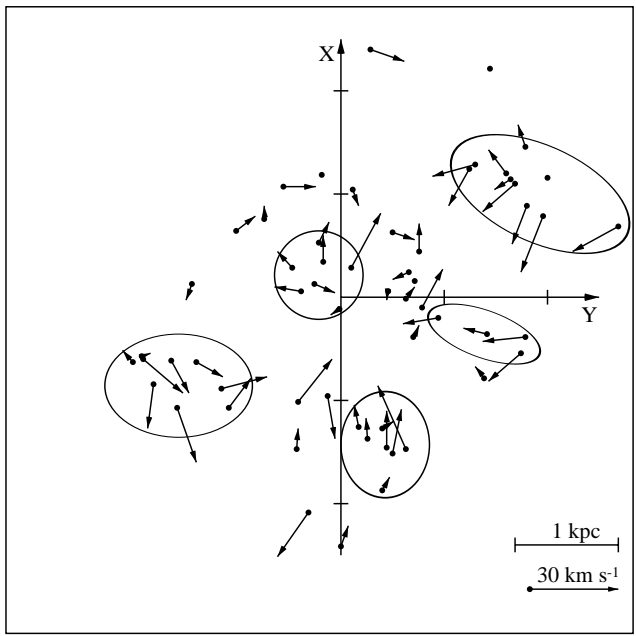}}
\caption{ The residual velocities of OB-associations derived for
the new reduction of the Hipparcos data and  $R_0=7.5$ kpc. The
ellipses indicate the positions of the stellar-gas complexes. The
$X$- and $Y$ axes are directed away from the Galactic center and
toward the Galactic rotation, respectively. The Sun is at the
origin.} \label{residual-velocities}
\end{figure}

The distributions of the residual velocities of OB-associations
derived for the old and new reductions of the Hipparcos data
resemble each other (compare Fig.~2 in \citet{melnik2001} and
Fig.~\ref{residual-velocities} of the present paper). The
velocities differ, on average, by  3.5 km s$^{-1}$.

Fig.~\ref{residual-velocities} also shows the grouping of
OB-associations into regions of intense  star formation, which
practically coincide with the stellar-gas complexes identified by
\citet{efremov1988}. For each complex we calculated the mean
residual velocities of OB-associations, which are listed in Table
3. Positive radial residual velocities $V_R$ are directed away
from the Galactic center, and the positive azimuthal residual
velocities $V_\theta$ are  in the sense of Galactic rotation.
Table 3 also contains the rms errors of the mean velocities, the
average Galactocentric distances $R$, the intervals of galactic
longitudes $l$, the intervals of heliocentric distances $r$, and
names of OB-associations the region includes.

Fig.~\ref{residual-velocities} and Table 3 clearly show that young
stars in some regions have conspicuous residual velocities. The
mean radial residual velocities in the Perseus, Cygnus, and Carina
regions are directed toward the Galactic center and are equal to
$V_R=-7$, $-5$, and $-6$ km s$^{-1}$, respectively, whereas those
in the Sagittarius region and in the Local System are directed
away from the center and are equal to $V_R=+10$ and $V_R=+5$ km
s$^{-1}$, respectively. As for the mean azimuthal residual
velocities, they are close to zero in the Sagittarius region and
in the Local System. In the Perseus and Cygnus regions the
direction of azimuthal motions is opposite that of Galactic
rotation and the corresponding velocity components are
$V_\theta=-6$ and $V_\theta=-10$ km s$^{-1}$, respectively,
whereas in the Carina region the azimuthal residual velocity,
which is equal to $V_\theta=+5$ km s$^{-1}$,  is directed in the
sense of Galactic rotation.

Similar kinematical features are observed in the old reduction of
the Hipparcos catalog. The mean residual velocities in the
stellar-gas complexes derived for the old and new reductions
differ, on average, by 1.0 km s$^{-1}$. The maximal difference
(2.1 km s$^{-1}$) is observed in the Cygnus region (compare Table
1 in \citet{melnik2009} with Table 3 of the present paper). Thus
the mean residual velocities  in the stellar-gas complexes depend
only slightly on the choice of the data-reduction way.

\section{Conclusions}

Proper motions of OB-associations derived for the old and new
reductions differ, on average, by $\Delta_{\mu_l}=0.58$  and
$\Delta_{\mu_b}=0.62$ mas yr$^{-1}$. This translates into a mean
velocity difference of $\Delta_{tang}=4.6$ km s$^{-1}$, which does
not exceed the standard deviation of the velocities of
OB-associations from the rotation curve, $\sigma=7.2$ km s$^{-1}$.
Generally, the new reduction  brings a weak decrease in the
dispersions of stellar proper motions inside OB-associations.

An analysis of the line-of-sight velocities and proper motions of
OB-associations shows that the Galactic rotation curve is
practically flat within 3 kpc of the Sun and corresponds to a high
angular rotation velocity at the solar distance, $\Omega_0=31\pm1$
km s$^{-1}$ kpc$^{-1}$. This result does not depend on the way of
the data reduction.

The values of the distance-scale coefficient $k$, $r=kr_{BH}$,
derived  from the kinematics of OB-associations,  $k=0.79\pm0.1$,
and trigonometric parallaxes, $k=0.78\textrm{--}0.88$, suggest
that the distance scale of \citet{blahahumphreys1989}  should be
shortened by 10--20\%.

We calculated the parameters of the rotation curve,  the
components of the solar motion, and the distance-scale coefficient
for different values of $R_0$. The values of $\Omega_0$, $u_0$,
$v_0$, and $k$ are practically independent on the choice of $R_0$.
The parameters $\Omega'_0$ and $\Omega''_0$ vary conspicuously
with  $R_0$, but the  Oort constant $A=-0.5R_0\Omega'$ remains
practically unchanged, $A=17.3\textrm{--}17.9$ km s$^{-1}$
kpc$^{-1}$.

The  residual velocities of OB-associations derived for the old
and new reductions of the Hipparcos data differ, on average, by
3.5 km s$^{-1}$. The differences in residual velocities decrease
down to 1.0 km s$^{-1}$ if OB-associations are grouped into the
stellar-gas complexes. The mean residual velocities of
OB-associations in the stellar-gas complexes depend only slightly
on the choice of the data-reduction procedure.

This work  was partly supported by the Russian Foundation for
Basic Research (project nos.~08\mbox{-}02\mbox{-}00738 and
~07\mbox{-}02\mbox{-}00380) and the Council for the Program of
Support for Leading Scientific Schools (project no.
NSh-433.2008.2).


\section{Appendix}

The 6\% estimate of the accuracy of relative distances is
determined by two independent factors.  First, because of nonzero
sizes of associations even exact distances of their individual
members may differ from the mean distance of the association by up
to half its line-of-sight size $D_r$. We can estimate $D_r$ as
$D_r \sim 2 \times \delta r$, where $\delta r$ is an association
"radius" in the sky plane (we assume that associations have, on
the average, the same sizes in the direction of line of sight and
in the sky plane), and hence even  exact relative distances of
association members may scatter by up to $\varepsilon_1 \leq
\delta r/r$. Second, distance estimates for individual members are
not exact, but are themselves determined with errors. The distance
moduli of all member stars are determined with more or less the
same standard error $\sigma_{DM}$ and the contribution of these
errors to the standard error of the distance modulus of the
association is $\sigma= \sigma_{DM}/\sqrt{n_t}$. We assume that
$\sigma_{DM}=0.3^m$ \citep{melnik1995}. Given that $DM = 5 \times
log_{10} r + 10$, where $r$ is distance in kpc, the relative error
of the association distance is $\varepsilon_2=10^{\sigma/5}-1$.
The combined relative error which includes both the scatter due to
the nonzero size  and that due to random errors is
$\varepsilon^2=\varepsilon_1^2+\varepsilon_2^2$ and its mean value
for OB-associations from the list of \citet{blahahumphreys1989} is
6\%.

\addtocounter{table}{-3}
\begin{table*}
  \caption{Line-of-sight velocities and proper motions of OB-associations (beginning)}
  \begin{tabular}{lcccccccccccccccc}
  \hline
Association&$l$&$b$&$r$&$V_r$&$\sigma_{vr}$&$n_{vr}$&$\mu_{l1}$&$\sigma_{\mu
l1}$&$\mu_{b1}$& $\sigma_{\mu b1}$&$\mu_{l2}$&$\sigma_{\mu
l2}$&$\mu_{b2}$&$\sigma_{\mu
b2}$&$n_{\mu}$&$N_t$\\
\hline
Sgr OB5  &  0.0& -1.2&2.4&-15.0&19.0& 2&  0.1&1.7& 0.1&2.2& -0.4&1.2& -1.2&2.2& 3& 31\\
Sgr OB1  &  7.6& -0.8&1.3&-10.0&12.1&37& -1.6&1.5&-1.3&1.1& -1.1&0.9& -1.1&1.3&29& 66\\
Sgr OB7  & 10.7& -1.6&1.4& -6.1&17.1& 3&  0.0&0.3&-3.3&0.1& -0.9&0.3& -1.3&0.8& 2&  4\\
Sgr OB4  & 12.1& -1.0&1.9&  3.5&10.8& 9& -0.7&1.4&-0.8&2.5& -0.5&1.5& -1.7&1.8& 3& 15\\
Sgr OB6  & 14.2&  1.3&1.6& -7.3& 9.5& 4& -0.3&   &-5.8&   &  0.1&   & -6.2&   & 1&  5\\
Ser OB1  & 16.7&  0.1&1.5& -5.0&21.6&17& -0.7&1.2&-0.8&2.6& -0.8&1.6& -0.1&1.8&12& 43\\
Sct OB3  & 17.3& -0.7&1.3&  3.3&17.0& 8& -0.9&0.5&-0.6&1.0& -0.6&0.5& -1.5&1.1& 3& 10\\
Ser OB2  & 18.2&  1.6&1.6& -4.0&14.5& 7& -0.8&1.4&-1.4&0.9& -0.3&1.3&  0.0&1.4& 5& 18\\
Sct OB2  & 23.2& -0.5&1.6&-11.0&40.2& 6& -0.5&4.2&-0.6&1.6& -1.8&4.0& -0.1&2.4& 6& 13\\
Tr 35    & 28.0& -0.5&2.0& 31.0&    & 1& -3.2&   &-0.1&   & -3.7&   &  1.8&   & 1&  9\\
Coll 359 & 29.8& 12.6&0.2&     &    & 0& -7.1&   &-4.1&   & -7.8&   & -5.3&   & 1&  1\\
Vul OB1  & 60.4&  0.0&1.6&  3.1&14.7& 9& -4.7&1.7&-0.1&0.7& -5.1&1.8& -0.1&0.9& 8& 28\\
Vul OB4  & 60.6& -1.2&0.8& -2.9& 7.5& 3& -4.1&1.1&-1.9&1.2& -3.6&0.4& -1.9&1.2& 3&  9\\
Cyg OB3  & 72.8&  2.0&1.8& -9.5&11.5&30& -7.6&1.6&-1.2&1.3& -7.3&1.5& -1.3&1.2&18& 42\\
Cyg OB1  & 75.8&  1.1&1.5&-13.5&10.5&34& -6.2&1.1&-0.8&1.3& -5.7&1.1& -0.5&0.8&14& 71\\
Cyg OB9  & 77.8&  1.8&1.0&-19.5&13.4&10& -6.7&1.4&-1.4&0.8& -6.2&1.2& -1.4&0.9& 8& 32\\
Cyg OB8  & 77.9&  3.4&1.8&-21.0&11.0& 9& -6.2&0.9& 0.8&1.4& -6.3&0.7&  1.4&2.0&10& 21\\
Cyg OB2  & 80.3&  0.9&1.5&     &    & 0& -3.9&   &-1.5&   & -5.4&   & -2.9&   & 1& 15\\
Cyg OB4  & 82.7& -7.5&0.8& -4.9& 1.1& 2& -0.7&1.6&-1.3&1.5& -1.1&1.6& -1.1&1.2& 2&  2\\
Cyg OB7  & 89.0&  0.0&0.6& -9.4& 9.3&21& -3.2&3.8&-1.1&1.4& -2.6&3.8& -1.3&1.4&28& 29\\
NGC 6991 & 87.6&  1.4&1.4&-15.0&    & 1& -4.7&   &-0.3&   & -4.3&   & -0.9&   & 1&  1\\
Lac OB1  & 96.7&-17.7&0.5&-13.6& 4.2& 2& -3.3&0.2&-4.4&0.4& -3.1&0.1& -4.5&0.1& 2&  2\\
Cep OB2  &102.1&  4.6&0.7&-17.0& 6.7&37& -3.8&1.5&-0.8&1.6& -3.7&1.4& -0.8&1.5&47& 59\\
Cep OB1  &104.2& -1.0&2.8&-58.2&14.0&17& -3.9&1.1&-0.4&1.0& -4.5&1.3& -0.7&1.0&24& 58\\
NGC 7235 &102.8&  0.8&3.2&     &    & 0&     &   &    &   &     &   &     &   & 0&  1\\
Cep OB5  &108.5& -2.7&1.7&-48.7&29.2& 2& -5.2&   &-0.4&   & -4.9&   &  0.1&   & 1&  6\\
Cas OB2  &112.0&  0.0&2.1&-50.1&11.0& 7& -4.5&2.1&-0.1&2.8& -4.8&1.8&  0.1&2.6& 5& 41\\
Cep OB3  &110.4&  2.6&0.7&-22.9& 5.0&18& -2.2&1.3&-1.7&1.6& -1.9&1.5& -2.0&1.8&15& 26\\
Cas OB5  &116.1& -0.5&2.0&-45.8&11.1&16& -3.6&1.1&-2.0&1.2& -3.8&1.2& -1.4&0.8&13& 52\\
Cep OB4  &118.3&  5.3&0.7&-24.0&    & 1& -2.2&1.2&-1.1&0.9& -2.3&0.7& -1.2&1.0& 4&  7\\
Cas OB4  &120.1& -0.3&2.3&-37.0& 8.6& 7& -2.1&1.2&-1.3&1.1& -2.5&1.0& -1.1&0.7& 7& 27\\
Cas OB14 &120.4&  0.7&0.9&-15.0&34.2& 4&  1.6&2.0&-1.1&1.1&  0.9&1.6& -2.3&1.2& 3&  8\\
Cas OB7  &123.0&  1.2&2.0&-50.0&16.1& 4& -2.0&0.7&-0.9&0.7& -2.3&0.5& -0.4&0.5& 8& 39\\
Cas OB1  &124.7& -1.7&2.0&-42.0&20.5& 5& -2.0&0.0&-1.6&1.4& -1.8&0.3& -1.7&1.4& 3& 11\\
NGC 457  &126.7& -4.4&2.0&-34.8& 9.8& 4& -1.6&0.2&-1.7&0.2& -0.5&0.2& -1.1&0.3& 2&  4\\
Cas OB8  &129.2& -1.1&2.3&-34.7& 9.9&14& -1.4&0.6&-1.0&1.9& -0.8&0.7& -0.6&1.2& 9& 43\\
Per OB1  &134.7& -3.2&1.8&-43.4& 7.0&81&  0.1&1.0&-1.7&1.4&  0.2&1.3& -1.5&1.2&63&167\\
Cas OB6  &135.0&  0.8&1.8&-42.7&10.0&12& -0.6&1.3&-1.2&1.4& -1.0&1.8& -1.1&1.6&13& 46\\
Cam OB1  &141.1&  0.9&0.8&-11.0&11.2&30&  0.6&1.9&-1.7&1.6&  0.0&2.0& -1.4&1.4&33& 50\\
Cam OB3  &147.0&  2.8&2.6&-27.6&19.3& 3&  1.6&2.0& 0.9&0.8& -0.1&1.5&  0.2&1.1& 3&  8\\
Per OB3  &146.6& -5.9&0.1& -2.4&    & 1& 34.5&   &-8.2&   & 34.3&   & -8.7&   & 1&  1\\
Per OB2  &160.3&-16.5&0.3& 21.2& 4.5& 7&  6.5&2.4&-1.3&1.7&  6.6&2.7& -2.8&2.4& 7&  7\\
Aur OB1  &173.9&  0.1&1.1& -1.9&14.0&26&  3.0&1.3&-2.0&1.9&  2.5&1.4& -1.9&1.3&20& 36\\
Ori OB1  &206.9&-17.7&0.4& 25.4& 7.9&62&  0.7&1.6& 0.6&2.0&  0.9&1.8&  0.3&1.9&59& 70\\
Aur OB2  &173.3& -0.2&2.4& -2.7& 5.5& 4&  0.0&1.0&-0.9&0.1& -0.1&0.3& -1.4&1.0& 2& 20\\
NGC 1893 &173.6& -1.7&2.9&     &    & 0&     &   &    &   &     &   &     &   & 0& 10\\
NGC 2129 &186.5& -0.1&1.5& 16.8& 7.5& 3&  2.7&   & 1.1&   &  2.7&   & -0.2&   & 1&  3\\
Gem OB1  &189.0&  2.2&1.2& 16.0& 6.0&18&  1.5&1.7&-1.1&1.5&  1.9&1.2& -0.9&1.3&17& 40\\
Mon OB1  &202.1&  1.1&0.6& 23.4&13.0& 7&  1.9&1.9&-2.1&2.2&  0.7&1.2& -2.4&1.2& 7&  7\\
Mon OB2  &207.5& -1.6&1.2& 22.3&12.5&26& -1.0&2.3&-1.2&1.6& -1.1&2.1& -1.8&1.5&18& 32\\
Mon OB3  &217.6& -0.4&2.4& 27.0&    & 1&     &   &    &   &     &   &     &   & 0&  4\\
CMa OB1  &224.6& -1.6&1.1& 34.3&16.1& 8& -3.2&1.1&-2.6&2.0& -2.3&1.6& -1.9&1.0&10& 17\\
NGC 2414 &231.1&  1.0&3.2& 67.2&    & 1& -2.6&   &-0.4&   & -2.0&   & -0.4&   & 1& 15\\
Coll 121 &238.5& -8.4&0.6& 29.6& 9.3&10& -5.2&1.2&-1.0&0.9& -5.1&1.3& -1.3&0.6&13& 13\\
NGC 2362 &237.9& -5.9&1.2& 30.0&16.8& 6& -4.2&1.5& 0.1&0.9& -5.1&0.9& -0.5&0.7& 3&  9\\
NGC 2367 &235.7& -3.8&2.2& 37.0&21.6& 4& -0.7&   & 0.1&   & -4.5&   &  1.4&   & 1&  5\\
NGC 2439 &245.3& -4.1&3.5& 62.7&    & 1& -4.5&1.0&-0.7&1.0& -4.7&1.1& -0.6&1.0&10& 23\\
Pup OB1  &243.5&  0.2&2.0& 77.0&    & 1& -3.6&1.4& 0.3&1.5& -3.8&1.7& -0.2&1.8& 4& 22\\
Pup OB2  &244.6&  0.6&3.2&     &    & 0&     &   &    &   &     &   &     &   & 0& 13\\
Coll 140 &244.5& -7.3&0.3& 10.3&10.5& 5& -7.0&1.8&-5.3&3.6& -6.8&1.5& -4.7&3.5& 6&  6\\
\hline
\end{tabular}
\end{table*}

\addtocounter{table}{-1}

\begin{table*}
\caption{Line-of-sight velocities and proper motions of
OB-associations (end)}
  \begin{tabular}{lcccccccccccccccc}
  \hline
Association&$l$&$b$&$r$&$V_r$&$\sigma_{vr}$&$n_{vr}$&$\mu_{l1}$&$\sigma_{\mu
l1}$&$\mu_{b1}$&
$\sigma_{\mu b1}$&$\mu_{l2}$&$\sigma_{\mu l2}$&$\mu_{b2}$&$\sigma_{\mu b2}$&$n_{\mu}$&$N_t$\\
\hline
Pup OB3  &253.9& -0.3&1.5&     &    & 0&     &   &    &   &     &   &     &   & 0&  3\\
Vela OB2 &262.1& -8.5&0.4& 24.0& 9.7&13& -9.7&3.2&-1.4&2.7& -9.5&2.7& -0.8&2.2&12& 13\\
Vela OB1 &264.9& -1.4&1.5& 23.0& 6.6&18& -6.7&2.1&-1.3&1.1& -6.2&2.0& -1.6&0.8&18& 46\\
Car OB1  &286.5& -0.5&2.0& -5.0& 8.2&39& -7.5&0.6&-1.0&0.5& -7.6&0.8& -0.6&1.1&18&126\\
Tr 16    &287.3& -0.3&2.1& -1.0& 8.9& 5& -7.2&0.5&-0.9&0.1& -8.0&0.4& -1.3&0.1& 2& 18\\
Tr 14    &287.4& -0.5&2.8&-10.0&    & 1&     &   &    &   &     &   &     &   & 0&  5\\
Tr 15    &287.6& -0.4&3.0&     &    & 0&     &   &    &   &     &   &     &   & 0&  2\\
Coll 228 &287.6& -1.0&2.0&-13.0& 9.0& 9& -6.7&0.7&-1.0&0.1& -6.0&1.0& -1.0&0.1& 2& 15\\
Car OB2  &290.4&  0.1&1.8& -8.2& 8.9&22& -6.8&1.1&-0.8&1.1& -6.1&1.1& -0.5&1.4&12& 59\\
NGC 3576 &291.3& -0.6&2.5&-17.0& 9.0& 2&     &   &    &   &     &   &     &   & 0&  5\\
Cru OB1  &294.9& -1.1&2.0& -5.3& 8.9&33& -6.0&1.0&-0.8&0.5& -6.0&0.7& -0.7&0.8&19& 76\\
NGC 3766 &294.1& -0.0&1.5&-15.6& 0.7& 2& -6.3&0.6&-0.9&0.1& -6.5&0.5& -1.1&0.1& 2& 12\\
Cen OB1  &304.2&  1.4&1.9&-19.0&14.5&32& -4.9&0.7&-0.9&0.9& -4.6&1.2& -0.9&1.0&32&103\\
Hogg 16  &307.5&  1.4&1.5&-35.0& 8.5& 3& -4.1&1.4&-2.6&2.9& -5.6&5.2& -2.3&3.8& 3&  5\\
R 80     &309.4& -0.4&2.9&-38.2&    & 1& -6.2&   &-2.4&   & -3.4&   & -1.9&   & 1&  2\\
NGC 5606 &314.9&  1.0&1.5&-37.8& 1.8& 3& -4.9&2.0&-1.7&0.4& -5.8&1.5& -2.3&0.4& 2&  5\\
Cir OB1  &315.5& -2.8&2.0&     &    & 0&     &   &    &   &     &   &     &   & 0&  4\\
Pis 20   &320.4& -1.5&3.2&-49.0&    & 1&-11.7&   & 1.4&   & -9.8&   &  2.6&   & 1&  6\\
Nor OB1  &328.0& -0.9&2.8&-35.6&22.5& 6& -3.2&   & 0.7&   & -4.1&   & -0.1&   & 1&  8\\
NGC 6067 &329.7& -2.2&1.7&-40.0& 2.6& 8& -2.3&   &-1.2&   & -2.0&   & -1.4&   & 1&  9\\
R 103    &332.4& -0.8&3.2&-48.0&30.5&11& -3.9&1.5&-2.0&1.8& -4.0&1.2& -2.5&2.3& 4& 34\\
R 105    &333.1&  1.9&1.3&-31.0&11.0& 4& -4.4&   &-1.3&   & -5.3&   & -2.8&   & 1&  4\\
Ara OB1B &338.0& -0.9&2.8&-34.7&10.4& 9& -3.6&2.1&-2.2&4.3& -3.0&2.9& -2.4&3.1& 6& 21\\
Ara OB1A &337.7& -0.9&1.1&-36.3&20.6& 8& -2.5&4.2&-2.2&2.3& -2.0&2.6& -2.0&3.2&10& 53\\
NGC 6204 &338.3& -1.1&2.2&-51.0&27.0& 5&     &   &    &   &     &   &     &   & 0& 14\\
Sco OB1  &343.7&  1.4&1.5&-28.8&17.5&28& -1.5&1.1&-0.7&1.2& -2.0&1.3& -1.3&0.9&16& 76\\
Sco OB2  &351.3& 19.0&0.1& -4.1& 4.8&10&-24.0&3.4&-8.7&5.0&-23.5&2.9&-10.5&4.0&10& 10\\
HD 156154&351.3&  1.4&2.1& -4.0& 8.5& 3& -1.2&0.1&-0.9&1.2& -1.7&0.2& -0.2&1.5& 2&  4\\
Sco OB4  &352.7&  3.2&1.0&  3.0& 6.3& 7& -0.8&1.3&-2.3&1.0& -0.9&0.8& -2.3&0.7& 4& 11\\
Tr 27    &355.1& -0.7&0.9&-15.8&    & 1&     &   &    &   &     &   &     &   & 0& 11\\
M 6      &356.8& -0.9&0.4& -6.4&    & 1& -6.6&   &-1.7&   & -6.6&   & -1.7&   & 1&  1\\
\hline
\end{tabular}
\end{table*}

\end{document}